\documentclass[fleqn,twoside]{article}
\usepackage{espcrc2,times}
\usepackage{graphicx}
\newcommand{\smcaption}[1]{\vspace*{-1cm}\caption{#1}\vspace*{-0.5cm}}
\newcommand{\beq}{\begin{equation}}
\newcommand{\eeq}{\end{equation}}
\newcommand{\bea}{\begin{eqnarray}}
\newcommand{\eea}{\end{eqnarray}}

\renewcommand{\d}{\delta}
\renewcommand{\l}{\lambda}
\renewcommand{\L}{\Lambda}
\renewcommand{\b}{\beta}

\newcommand{\n}{\nu}
\newcommand{\m}{\mu}
\renewcommand{\r}{\rho}

\newcommand{\s}{\sigma}

\newcommand{\E}{{\cal E}}

\newcommand{\tU}{\widetilde{U}}

\newcommand{\oh}{{\textstyle{\frac{1}{2}}}}

\newcommand{\non}{\nonumber}

\newcommand{\rf}[1]{(\ref{#1})}
\newcommand{\ra}{\rightarrow}
\newcommand{\pa}{\partial}

\newcommand{\AmS}{{\protect\the\textfont2
  A\kern-.1667em\lower.5ex\hbox{M}\kern-.125emS}}

\title{Confinement and center vortices in Coulomb gauge: analytic and numerical results\thanks{Talk 
given by D. Zwanziger at {\it QCD Down Under}, Adelaide, Australia, March 10--19, 2004.
Research is supported in part by the U.S. Department of Energy
under Grant No.\ DE-FG03-92ER40711 (J.G.), the Slovak Grant Agency for Science, Grant 
No. 2/3106/2003 (\v{S}.O.), and the National Science Foundation, Grant No. PHY-0099393 (D.Z.).}}

\author{Jeff Greensite\address{The Niels Bohr Institute,
Blegdamsvej 17, DK--2100 Copenhagen \O, Denmark}, 
{\v S}tefan Olejn\'{\i}k\address{Institute of Physics, Slovak Academy
of Sciences, SK--845 11 Bratislava, Slovakia}, 
and Daniel Zwanziger\address{Physics Department, New York
University, New York, NY~10003, USA} }

\begin{document}
\begin{abstract}

   We review the confinement scenario in Coulomb gauge.  We show that when 
thin center vortex configurations are gauge transformed to Coulomb gauge, they 
lie on the common boundary of the fundamental modular region and the Gribov region.  
This unifies elements of the Gribov confinement scenario in Coulomb gauge and the center-vortex 
confinement scenario.  We report on recent numerical studies which support both of these scenarios. 
     
\end{abstract}
\maketitle
%
% Section I
%
\section{Introduction}\label{introduction}

	In a confining theory such as QCD it is helpful to choose a gauge which makes the confining 
mechanism transparent.  In the present article we shall be primarily concerned with the 
confinement scenarios in minimal Coulomb gauge and in maximal center gauge, and 
with the unification of elements of these two scenarios.  
	
	The Landau gauge has the simplest Lorentz transformation properties, 
but there is at present no confinement scenario in Landau gauge.  The obstacle is that 
the gluon propagator is of shorter range in Landau gauge than it is in a free theory, because of the 
suppression of the low-momentum components by the proximity of the Gribov horizon (see below).  
This makes the mechanism of confinement of color charge in Landau gauge more rather than less 
mysterious.  We turn instead to the Coulomb gauge.
	
	 Confinement of color charge is easily understood in minimal Coulomb gauge because 
the 0-0 component of the gluon propagator, 
\beq
D_{00}(x,t) = V_{\rm coul}(|x|) \ \d(t) + \mbox{non-instantaneous},
\label{timeprop}
\eeq
has an instantaneous part, $V_{\rm coul}(r)$, that is long range and confining and 
couples universally to all color-charge.
We call $V_{\rm coul}(r)$ ``the color-Coulomb potential''.  Moreover the
3-dimensionally transverse would-be physical components of the gluon propagator,
\beq
D_{ij}(x,t) = \langle A_i(x,t) A_j(0,0) \rangle,
\label{spaceprop}
\eeq 
are short range, corresponding to the absence of gluons from the physical spectrum.   
We shall review recent numerical evidence  \cite{JS,Us1,Us2} which strongly supports 
these statements.

The Coulomb gauge has also been studied vigorously in the hamiltonian formalism \cite{szcz}.

\section{Definition of minimal Coulomb gauge}\label{definition}

	In minimal Coulomb gauge, the representative $A_i(x)$ of each gauge orbit (at a fixed time~$t$) is the absolute minimum of a 
minimizing functional with respect to gauge transformations.  The minimizing functional is taken to be
\beq
F_A(g) \equiv ||{^g}A||^2
\eeq  
where the gauge transform is given by
\beq
{^g}A_i = g^{-1}A_i g + g^{-1} \pa_i g,
\eeq
and the Hilbert norm by,
\beq
||A||^2 = \int d^3x \sum_i^a |A_i^a(x)|^2.
\eeq
The set of representatives,   $\Lambda$, satisfies
\beq
\L \equiv \{A_i(x): ||A|| \leq ||{^g}A|| \},
\eeq
for all gauge transformations $g(x)$, and is called the ``fundamental modular region''.

     In lattice gauge theory, the minimal Coulomb gauge is defined analogously. A maximizing functional is defined by
\beq
F_U(g) = \sum_{x,i} {\rm Re \ Tr} [ {^g}U_{x,i} ]
\eeq
and the fundamental modular region by
\beq
\L = \{U: F_U(1) \geq F_U(g) \}.
\eeq
The maximization is done independently within each time slice $t$.   
Euclidean and Minkowski Coulomb gauges are the same.  
Only physical excitations are propagated in the time direction. 

In practice it is difficult to find the absolute minimum, and one is satisfied to find a relative minimum.  In principle one should check the sensitivity to choice of minimum.

\section{Elementary properties of minimal Coulomb gauge}\label{elementary}

\begin{figure}[t!]
%I percented out the followling line
%\centering\includegraphics[width=6truecm]{orbitspace.pdf}
\centerline{\includegraphics[width=6truecm]{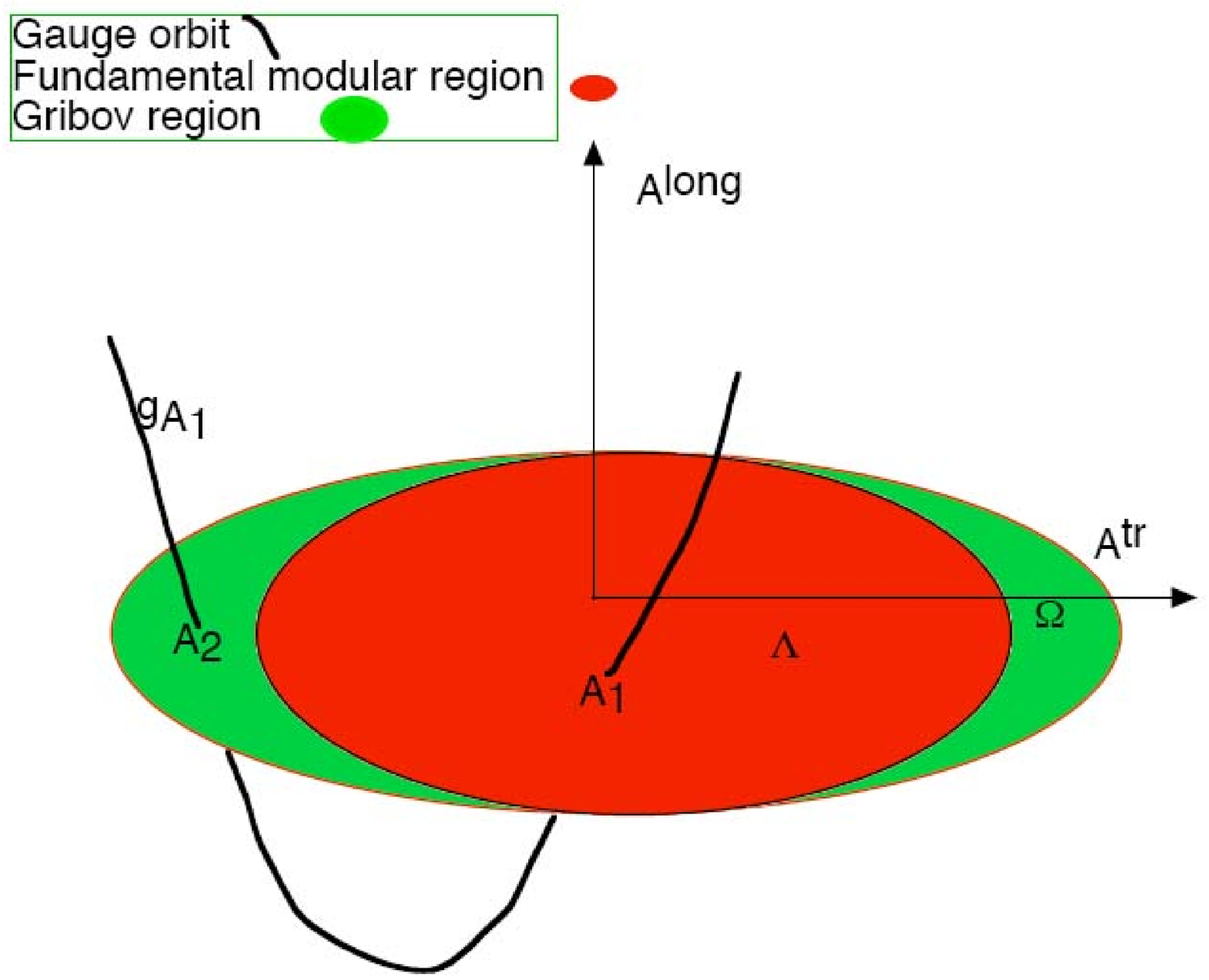}}
\smcaption{A typical gauge orbit through the configuration $A_1$ is represented as a curve.  Transverse configurations lie in the horizontal plane, viewed from above.  It contains $\Omega$ and $\Lambda$, and the pair of Gribov copies $A_1$ and $A_2$.}
\label{o}
\end{figure}

     At a relative or absolute minimum, 
the minimizing functional is stationary, which yields the transversality or Coulomb-gauge condition
\beq
\pa_i A_i(x) = 0.
\label{transversality}
\eeq
In addition, the matrix of second derivatives is non-negative.  In the present case it is the Faddeev--Popov operator,
\beq
M(A) \equiv - \pa_i D_i(A) \geq 0,
\label{fadpop}
\eeq
where $
D_i^{ac}(A) = \pa_i \d^{ac} + f^{abc} A_i^b(x)$ 
is the gauge-covariant derivative.  These two properties hold at any relative or absolute minimum.  Together they define the Gribov region
\beq
\Omega = \{A: {\rm properties \ \rf{transversality} \ and \ \rf{fadpop} \ hold} \}.
\eeq
The set of relative minima  $\Omega$  is larger than the set of absolute minima $\L$, and we have the inclusion
$\L \subset \Omega$,
%\eeq
as illustrated in Fig.~\ref{o}.  
In continuum gauge theory both $\L$ and $\Omega$ are convex and bounded in every direction \cite{semenov},
and in SU(2) lattice gauge theory a slightly weaker convexity property is established in \cite{Us2}.

\section{ No confinement without Coulomb confinement}\label{noconfine}

   The gauge-invariant, physical potential $V(R)$ between a pair of external quarks is obtained from a rectangular Wilson loop, 
of dimension $R \times T$, with $T$ large,
\beq
    \exp[ - V(R) T ] = \langle  P \exp \int A_\m dx^\m \rangle .
\eeq
Whereas 
$V(R)$ involves $n$-point functions of all orders, the color-Coulomb potential $V_{\rm coul}(R)$, defined in \rf{timeprop}, is 
the instantaneous part of $D_{00}$, a 2-point function.  

     The ground state energy lies lower than the Coulomb energy, 
which leads to the interesting inequality \cite{Dan}
\beq
	E_{\rm se} + V(R) \leq E'_{\rm se} + V_{\rm coul}(R),
\eeq
where $E_{se}$ and $E'_{se}$ are self-energies that are finite in the presence of a lattice cut-off.  This bound is obtained from a trial wave-functional, and relies on the fact that the same kernel
$K_{xy}^{ab}(A)$
whose expectation-value is the color-Coulomb potential,
\beq
	 V_{\rm coul}(|x - y|) \d_{ab} = \langle  K_{xy}^{ab}(A) \rangle, 
\label{kernelofv}
\eeq
also appears in the Hamiltonian in Coulomb gauge for a static quark pair at $x$ and $y$,
\bea
	H &=& H_{\rm glue} + H_{\rm quark}
\non \\
(\Phi, H_{\rm quark} \Psi)  
    &=&  ((t^a)^* \Phi,  K_{xy}^{ab}(A) t^b  \Psi)
   + \dots
\eea
This kernel has the explicit form
\beq 
	K_{xy}^{ab}(A) \equiv 
[ M^{-1}(A) (- \nabla^2) M^{-1}(A) ]_{xy}^{ab}, 
\label{kernel}
\eeq
where $M(A)$ is the Faddeev--Popov operator \rf{fadpop}.

It follows that if $V(R)$ is confining,
\beq
	\lim_{R \ra \infty} V(R) = \infty,
\eeq
then the 
color-Coulomb potential, $V_{\rm coul}(R)$, is confining,
\beq
	\lim_{R \ra \infty} V_{\rm coul}(R) = \infty.
\eeq
Moreover if both 
increase linearly at large $R$,
$V(R) \sim \s \ R; \ V_{\rm coul}(R) \sim \s_{\rm coul} \ R$,
then
\beq
	\s_{\rm coul} \geq \s.
\eeq
We conclude that in the confining phase the 2-point function $V_{\rm coul}(R)$ is confining.

	Greensite and Olejn{\'\i}k \cite{JS} determined the color-Coulomb potential $V_{\rm coul}(R)$ numerically from the lattice quantity 
\bea
      G(|x|) & \equiv & \oh \langle {\rm Tr} [ U_0^\dag(x, t) U_0(0, t) ]  \rangle ,
\non \\
       V_{\rm coul}(R) & = & - \ln G(R).
\label{colorcoul}       
\eea
They found an impressively linear behavior of the color-Coulomb potential $V_{\rm coul}(R)$.  The relevant figure is presented in 
%Jeff 
Greensite's talk at this meeting~\cite{greensite}.

\section{Confinement scenario in Coulomb gauge}\label{scenario}

	To see why $V_{\rm coul}(R)$ is long range, consider formula~\rf{kernel} for the kernel $K_{xy}^{ab}(A)$.  Recall that  $M(A)$ is strictly positive in the interior of the Gribov region $\Omega$,  and develops a zero eigenvalue on its boundary (the Gribov horizon),
\bea
	M(A) & > & 0  \ {\rm for} \ A \ {\rm inside} \ \Omega, 
\non \\
	M(A) \phi_0 & = & 0 \ {\rm for} \  A \ {\rm on} \ \pa \Omega.
\eea
By continuity, $M(A)$ has a small eigenvalue for configurations near the boundary.  As explained below, the population is concentrated close to the boundary $\pa \Omega$.  This enhances $M^{-1}(A)$, and thus also $K_{xy}^{ab}(A)$, which makes $V_{\rm coul}(R)$ long range.  A more sophisticated discussion, given below, explains the long range of $V_{\rm coul}(R)$ in the infinite-volume limit in terms of the density of states $\r(\l)$ of the Faddeev--Popov operator $M(A)$.

 As already noted by Gribov~\cite{horizon}, the Gribov horizon is close by in directions (in $A$-space) that correspond to low-momentum Fourier components of the gluon field $A$.  This suppresses the low-momentum parts of {\it all} Lorentz components of the gluon propagator $D_{\m\n}$ in minimal Landau gauge.  While this eliminates gluons from the physical spectrum, it makes it harder to explain confinement of colored quarks.  For the same reason, in minimal Coulomb gauge, restriction to the interior of the Gribov horizon {\it suppresses} the low-momentum components of the would-be physical, 3-dimensional gluon propagator $D_{ij}(x,t)$,
which eliminates physical gluons from the physical spectrum.
However, as we have just seen, in Coulomb gauge it also {\it enhances} $V_{\rm coul}(|x|)$, which is the instantaneous part of $D_{00}$ in Coulomb gauge.  This couples universally to all colored objects, and is a prime candidate to trigger confinement.  Thus we can have our cake and eat it:  physical gluons are suppressed 
because $D_{ij}$ is short range, while $D_{00}$ is long range and can cause confinement.

\section{Double boundary dominance}\label{doubledom}

	The dimension $D$ of $A$-space in the presence of the lattice cut-off is a large number, of the order of the volume of the lattice $D \sim L^4$ , and in a space of high dimension the volume density goes like $r^{D-1}dr$.  We thus expect on entropy grounds that 
(a)~the population is concentrated close to the boundary $\pa \L$.  
However, as we have just seen, the enhancement of $V_{\rm coul}(R)$ occurs if (b)~the population is concentrated close to the boundary $\pa \Omega$.  However $\L$ is a subset of $\Omega$, 
$ \L \subset \Omega $, 
so (a) and (b) are compatible only if (c)~the population is concentrated where the 2 boundaries coincide,
%\beq 
i.e.\ $\pa \L \cap \pa \Omega$ .
%\eeq

Why should this be?  We present an explanation in terms of dominance of center vortices, and thereby unify elements of the Gribov confinement scenario in Coulomb gauge and the center-vortex confinement scenario.  The latter is reviewed in~\cite{review}.

\section{ Thin center vortex configurations lie on the double  boundary}\label{vortexondouble}

	Definition:  we call a ``thin center vortex configuration'' any lattice configuration for which every link variable is a center element, $U_{x,i} = Z_{x,i}$.  The unification of elements of the Gribov and center-vortex scenarios follows from the fact \cite{Us1} that when a center configuration $Z$ is gauge transformed to the minimal Coulomb gauge 
%\beq
$Z \ra U = {^g}Z \in \L$
%\eeq
it lies on the common boundary,
$U \in \pa
 \Omega \cap \pa \L$.
The proof is sketched below.  It follows that center vortex dominance implies dominance by a subset of this common boundary.

Note: The same statement and conclusion hold also for abelian configurations.

\section{Thin center vortex configurations lie on singular gauge orbits}\label{singularorbit}

Thin center vortex configurations and their gauge transforms $U = {^g}Z$ may be characterized in a gauge-covariant way as possessing non-zero solutions $\phi_n$
\beq
	D(U) \phi_n = 0,	                                      
\eeq
where $n = 1,\dots, N^2 -1$ for $SU(N)$, and $D(U)$ is the generator of a gauge transformation, defined by
\beq
{^g}U = U + \epsilon D(U) \phi,
\eeq
where $g = 1 + \epsilon \phi$ is an infinitesimal gauge transformation.  Indeed, center configurations are invariant,
${^h}Z = Z$, under the $N^2 - 1$ linearly independent global ($x$-independent) gauge transformations $h$.  For $h = 1 + \epsilon \phi$
this gives the gauge-covariant condition $D(Z)\phi = 0$.    

Note:  Similar properties hold for abelian configurations, with $n = 1,\dots, R$, where $R$ is the rank of the gauge group.
 
{\it Proof that a thin center vortex configuration lies on the double boundary:}
The equation $D(Z) \phi_n = 0$, being gauge covariant, holds after
transformation to minimal Coulomb gauge, 
$U = {^g}Z$, so 
$U \in \L$, and with $\psi_n = {^g}\phi_n$, we have $D(U) \psi_n = 0$.
It follows that the $\psi_n$ are null vectors of the Faddeev--Popov operator 
\beq
- \nabla_i D_i(U) \psi_n = 0,
\eeq
so $U = {^g}Z$ lies on the boundary of the Gribov region, 
$U \in \pa \Omega$.  We also have $U \in \Lambda$, and since 
$\L \subset \Omega$, it follows that $U$ lies on the double boundary, $U \in \pa \L \cap \pa \Omega$, as asserted.

The gauge orbit of a thin center configuration is a geometrically singular gauge orbit.  It has $N^2-1$, fewer dimensions than a generic gauge orbit, because 
${^g}U = U$ for $g(t) = \exp(t \phi_n)$.  By contrast, the Gribov horizon $\pa \Omega$ is, in general, merely a coordinate singularity.

\section{Numerical tests of center vortex dominance in Coulomb gauge}\label{numtest}

In \cite{Us2} the hypothesis of center vortex dominance was tested by the following procedure \cite{dFE}.  
(1) Configurations are fixed to the maximal center gauge.  (2)~In this gauge, a thin center-vortex configuration is defined, for SU(2), by
\beq
Z_{x,\m} = {\rm sign} [ {\rm Tr}(U_{x,\m}) ],
\eeq
and a ``vortex-removed" configuration by
\beq
\tU_{x,\m} = Z_{x,\m} U_{x,\m}.
\eeq
(3)  Finally, the resulting center-projected configuration $Z_{x,\m}$ and the vortex-removed configuration $\tU_{x,\m}$ are gauge transformed to the minimal Coulomb gauge.  According to the center vortex scenario, the projected center configurations should retain the confining properties, whereas the vortex-removed configurations should be non-confining.  

	The color-Coulomb potential $V_{\rm coul}(R)$, defined in 
\rf{colorcoul} was determined numerically in \cite{JS}, for both the full configuration and the vortex removed configuration.  The relevant figure is reproduced in Jeff Greensite's talk at this meeting~\cite{greensite}.  One sees that the color-Coulomb potential is impressively linear for the full configuration, as noted above, but is flat for the vortex-removed configurations.  This confirms the center-vortex scenario. 
	
	A more detailed test of the center-vortex scenario comes from a study \cite{Us2} of the density $\r(\l)$ of the eigenvalues $\l$ of the Faddeev--Popov operator \rf{fadpop}.  This quantity appears in the formula for the Coulomb or unscreened self-energy of a static color-charge at infinite lattice volume that follows from~\rf{kernelofv} and~\rf{kernel},
\beq
     \E = \int_0^{\l_{max}} {d\l  \over \l^2} \Bigl\langle \r(\l) F(\l) 
              \Bigr\rangle ,
\eeq
where
%\beq
    $ F(\l) \equiv \langle \l | (- \nabla^2) | \l \rangle $
%\eeq
is the diagonal matrix element of $(- \nabla^2)$ in the Faddeev--Popov eigenstates $| \l \rangle$.
The Coulomb energy is infrared divergent, as it should be in the confined phase, if, at infinite volume,
\beq
          \lim_{\l \ra 0} {\textstyle\frac{1}{\l}}{\left\langle \r(\l) F(\l) \right\rangle} > 0.
\label{condition}
\eeq
The quantities $\r(\l)$ and $F(\l)$ were determined numerically~\cite{Us2} for the full configurations, for center-projected (vortex-only) configurations and vortex-removed configurations, each of which have been transformed to Coulomb gauge.  

\begin{figure}[t!]
%I percented out the followling line
%\centering\includegraphics[width=6truecm]{r_2p1_all_full.pdf}
\centering\includegraphics[width=6truecm]{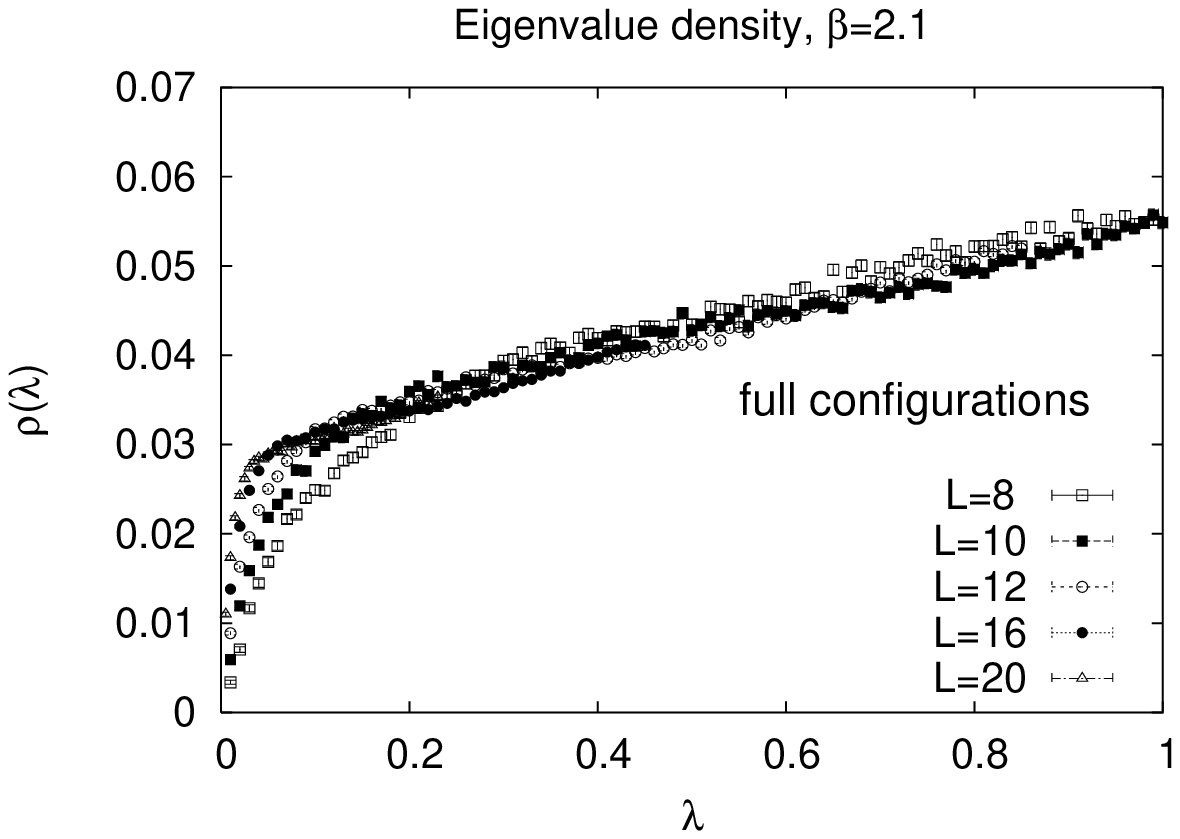}
\smcaption{The F-P eigenvalue density at $\b=2.1$, on
$8^4-20^4$ lattice volumes.}
\label{ra}
\end{figure}

\begin{figure}[t!]
%I percented out the followling line
%\centering\includegraphics[width=6truecm]{f_2p1_all_full.pdf}
\centering\includegraphics[width=6truecm]{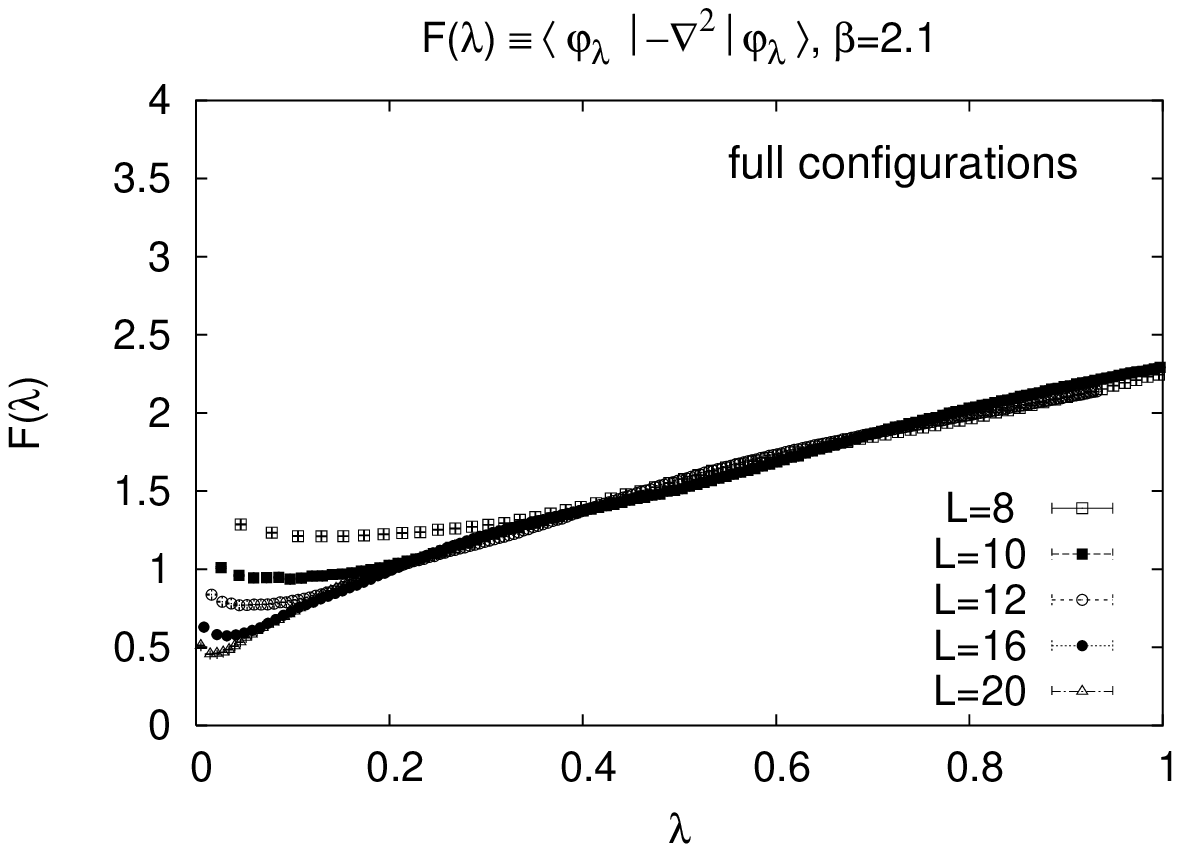}
\smcaption{$F(\l)$, the diagonal matrix element of $(-\nabla^2)$ 
in F-P eigenstates, plotted vs.\ F-P eigenvalue.}
\label{fa}
\end{figure}

Figures \ref{ra} and \ref{fa} show the results for $\langle \r(\l)\rangle$
and $\langle F(\l) \rangle$
for the full configurations, on a variety of lattice volumes
ranging from $8^4$ to $20^4$.  The apparent sharp ``bend" in
$\r(\l)$ near $\l=0$ becomes increasingly sharp, and happens ever
nearer $\l=0$, as the lattice volume increase. The impression
these graphs convey is that in the limit of infinite volume,
both $\rho(\l)$ and $F(\l)$ go to positive constants as $\l \ra 0$.
However, for both $\rho(\l)$ and $F(\l)$ we cannot 
exclude the possibility that the curves behave like $\l^p,~\l^q$ near
$\l=0$, with $p,q$ small powers.		

\begin{figure}[t!]
%I percented out the followling line
%\centering\includegraphics[width=6truecm]{r_2p1_all_cp.pdf}
\centering\includegraphics[width=6truecm]{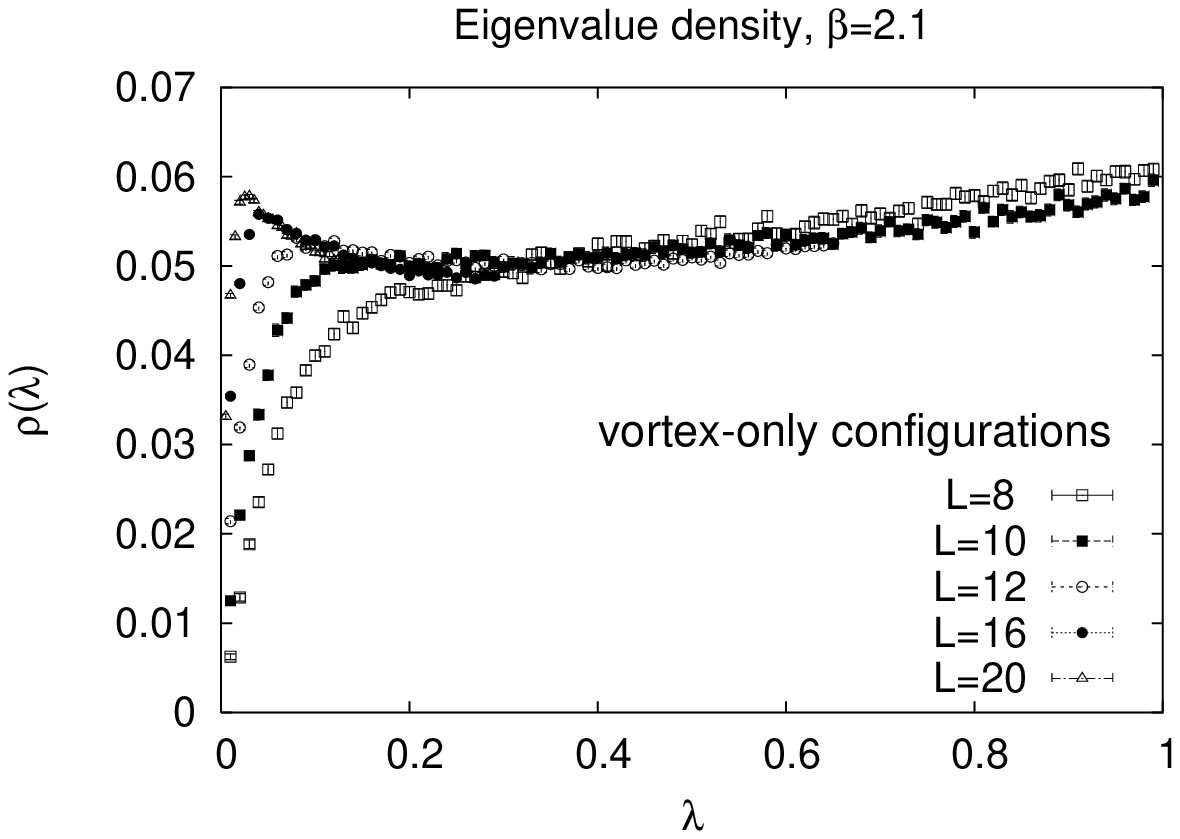}
\smcaption{F-P eigenvalue density in vortex-only configurations.}
\label{rcpa}
\end{figure}

\begin{figure}[t!]
%I percented out the followling line
%\centering\includegraphics[width=6truecm]{f_2p1_all_cp.pdf}
\centering\includegraphics[width=6truecm]{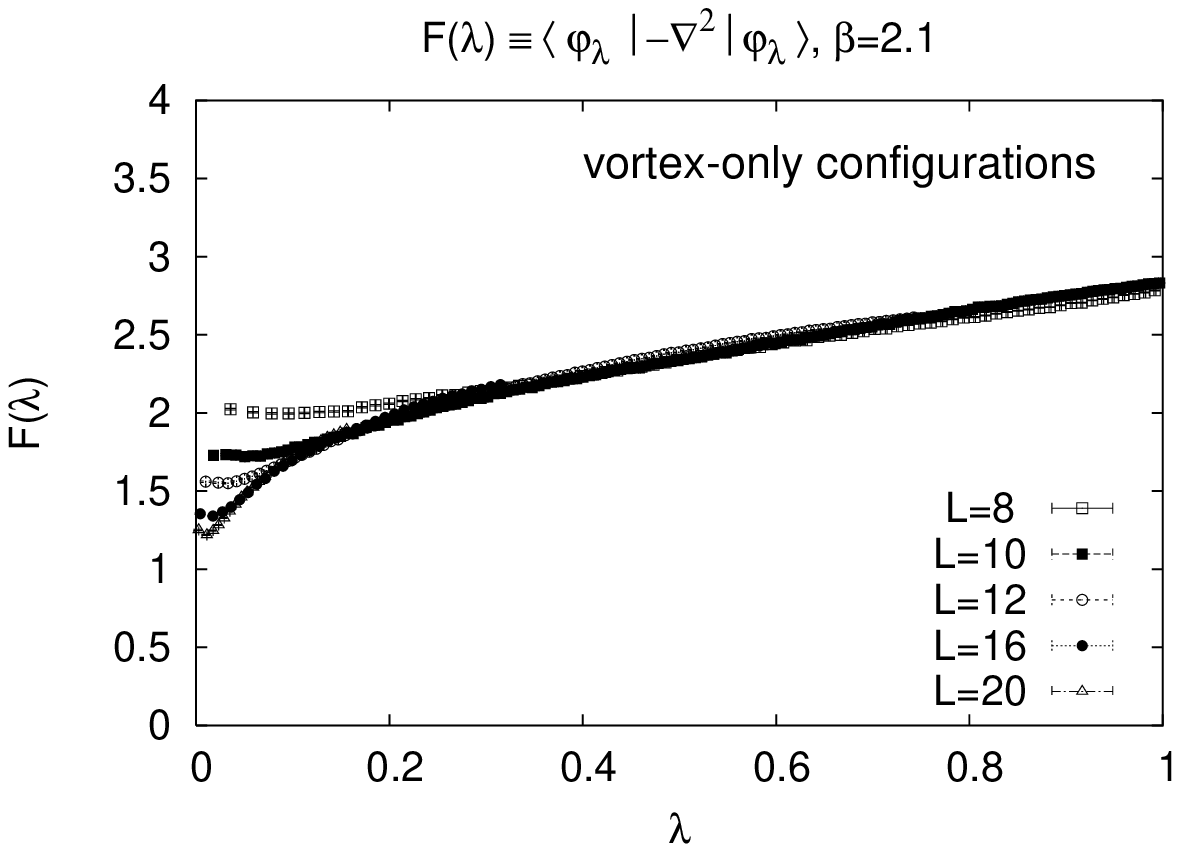}
\smcaption{$F(\l)$, the diagonal matrix element of $(-\nabla^2)$
in F-P eigenstates, for vortex-only configurations.}
\label{fcpa}
\end{figure}

   Next we consider the same observables for the ``vortex-only" configurations,
consisting of thin center vortex configurations transformed to Coulomb gauge.  The data for $\langle \r(\l)\rangle$ and $\langle F(\l)\rangle$ at the
same range ($8^4-20^4$) of lattice volumes are displayed in Figs.\
\ref{rcpa} and \ref{fcpa}.
The same qualitative features seen for
the full configurations, e.g.\ the sharp bend in the eigenvalue
density near $\l=0$, becoming sharper with increasing volume, are
present in the vortex-only data as well, and if anything are more
pronounced. 

\begin{figure}[t!]
%I percented out the followling line
%\centering\includegraphics[width=6truecm]{r_2p1_nv.pdf}
\centering\includegraphics[width=6truecm]{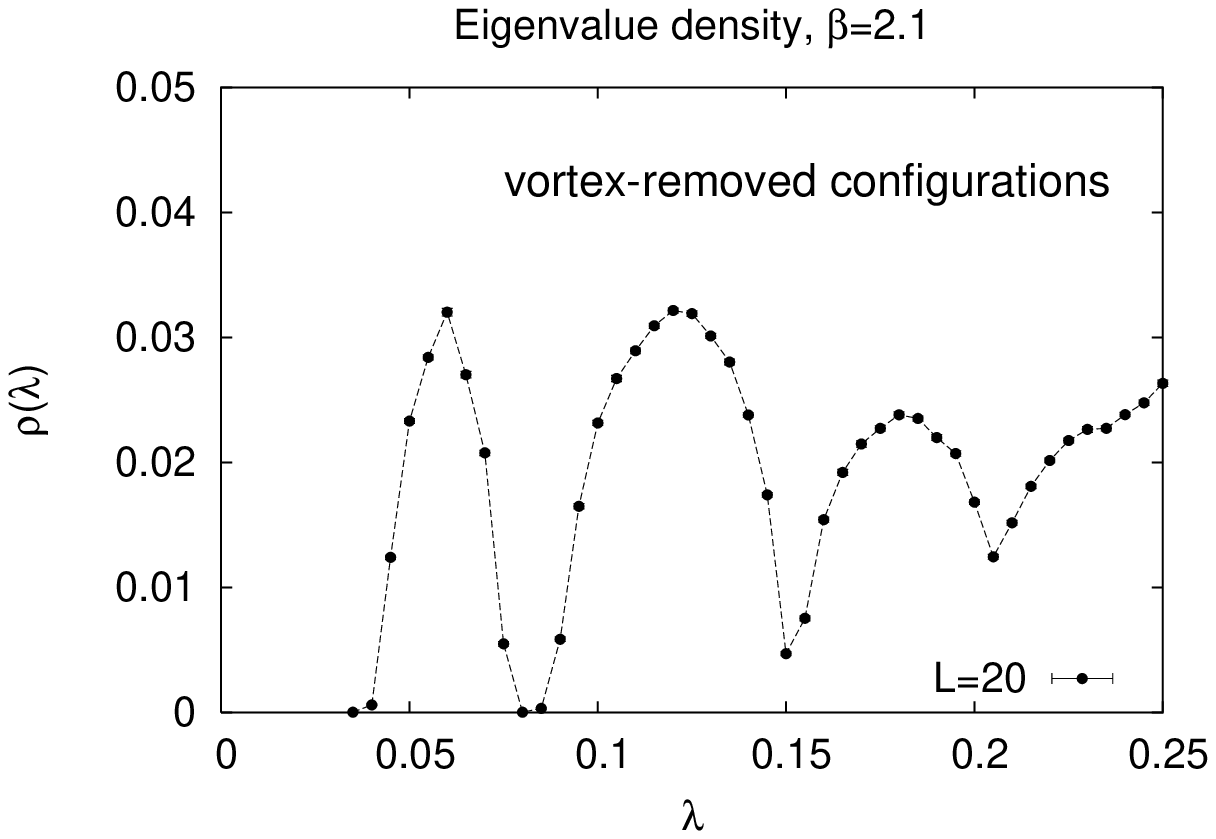}
\smcaption{F-P eigenvalue densities for vortex-removed
configurations, on a $20^4$ lattice volume.}
\label{rnva}
\end{figure}

   Finally, we consider the same observables for the vortex-removed
configurations, transformed to Coulomb gauge.  Results are
shown in Fig.\ \ref{rnva} for $\langle \r(\l)\rangle$.  
The behavior is strikingly
different, in the vortex-removed configurations, from what is seen in
the full and vortex-only configurations.  The graph of 
$\langle \r(\l)\rangle$, at each lattice volume, shows a set of distinct peaks, while the data for $\langle F(\l)\rangle$ (not shown) is organized into bands, with a slight gap between each band.  Inspection shows that eigenvalue
interval associated with each band in $\langle F(\l)\rangle$ precisely
matches the eigenvalue interval of one of the peaks in 
$\langle\r(\l)\rangle$.

   In order to understand these features, consider the eigenvalue
density of the Faddeev--Popov operator $M^{ab}_{xy}=\d^{ab}(-\nabla^2)_{xy}$
appropriate to an abelian theory, or a non-abelian theory at
zeroth order in the coupling.  At finite lattice volume, this operator has degenerate eigenvalues, and we call $N_k$ the degeneracy of its $k$-th eigenvalue $\l_k$.  When the degeneracies $N_k$, of the zeroth-order Faddeev--Popov eigenvalues are compared with the number of eigenvalues per lattice configuration found inside the 
$k$-th ``peak" of $\langle \r(\l)\rangle$, and $k$-th ``band" of $\langle F(\l)\rangle$, there is a precise match.  This leads to a simple interpretation:
the vortex-removed configuration $\tU_\m$ can be treated as a
small perturbation of the zero-field limit $U_\m=I$.  This
perturbation lifts the degeneracy of the $\l_k$, spreading the
degenerate eigenvalues into the peaks of finite width in
$\langle \r(\l)\rangle$ seen in Fig.~\ref{rnva}.  For the vortex-removed configurations, both 
$\langle \r(\l)\rangle$ and $\langle F(\l)\rangle$ seem to be only a perturbation of the corresponding zero-field results.

   We conclude that it is the vortex content of the thermalized
configurations which is responsible for the enhancement of both
$\r(\l)$ and $F(\l)$ near $\l=0$, leading to an
infrared-divergent Coulomb self-energy.

\section{Conclusion}\label{conclusion}

We  have seen that when thin center vortex configurations are gauge transformed to the minimal Coulomb gauge they lie on the double boundary of the Gribov region and the fundamental modular region.  This unifies elements of the Gribov confinement scenario in minimal Coulomb gauge and the center-vortex confinement scenario.  

Our numerical study \cite{Us2} reveals the following features:

(1)  The data are consistent with a linearly rising color-Coulomb potential, 
$V_{\rm coul}(R) \sim \s_{\rm coul} \ R$,  
and a Coulomb string tension that is larger than the physical string tension,
%\beq
$\s_{\rm coul} > \s$.
%\eeq

(2)  The data are compatible with a density of eigenvalues $\r(\l)$ of the Faddeev--Popov operator with either $\r(0)$ finite, or with $\r(\l) \sim \l^p$, for small $\l$, where $p$ is close to zero.  This feature, and the value of $\langle \rho(\l) F(\l) \rangle$ at small $\l$, provide detailed verification of the Gribov confinement scenario in Coulomb gauge.

(3)  These confining features are preserved by the ``vortex-only" configurations, but are
replaced by features close to a free theory in the ``vortex-removed" configurations.  This is consistent with center vortex dominance.  This, in turn, implies the condition of double boundary dominance that accords with the Gribov scenario in minimal Coulomb gauge.

Finally, we refer to \cite{Us2} for a similar numerical investigation of the gauge field coupled to a Higgs field, and of pure gauge theory at finite temperature in the deconfined phase.

%\bigskip
%
% Acknowledgements
%
%\acknowledgments{
%
%Our research is supported in part by }

%
% References
%

\end{document}